\documentclass[runningheads]{llncs}
\usepackage{graphicx}
\begin{document}
\title{Parallel Modeling of the Acoustic Signal Propagation in a Cased Well\thanks{Supported by Russian Science Foundation project 22-11-00104}}
\titlerunning{Acoustic signal in a cased well}
%
\author{Vladimir Cheverda \inst{1}\orcidID{0000-0002-5943-5901} \and
Galina Reshetova\inst{2}\orcidID{0000-0003-2653-3166} \and
Artem Kabannik\inst{3} \and Andrey Fedorov\inst{3} \and Roman Korkin\inst{3} \and Demid Demidov\inst{3} \and Viktor Balalayev\inst{3} }
\authorrunning{Cheverda, Reshetova, Kabannik, Fedorov, Korkin, Demidov, Balalayev}
%
\institute{Sobolev Institute of Mathematics, 630090, Novosibirsk, Russia \and Institute of Computational Matematics and Mathematical Geophysics \and Novosibirsk Technology Center Schlumberger \\
\email{vova\underline{\hspace{0.25cm}}chev@mail.ru}\\
\email{kgv@nmsf.sscc.ru}\\
\email{akabannik@slb.com}\\
\email{afedorov@slb.com}\\
\email{rkorkin@slb.com}\\
\email{ddemidov@slb.com}\\
\email{vbalalayev@slb.com}}
\maketitle              
\begin{abstract}
A numerical method is proposed for carrying out a full-scale simulation of the process of propagation of an acoustic signal in a cased well. The main goal is to study the interaction of the wave field with the vertical boundary of the cement filling the annular and near-pipe space. Particular attention is paid to the analysis of the intensity of the wave reflected from this boundary, depending on the degree of hardening of the upper edge of the cement.\\
A distinctive feature of the problem is the presence in it of several significantly different spatial scales. So, the length of the well can vary from several hundred meters to several kilometers, the diameter of the well is a few tens of centimeters, and the thickness of the casing string, as a rule, does not exceed a few first centimeters.
It is this diversity of scale that requires the organization of parallel computing, the organization of which is based on the spatial decomposition of the region and the involvement of Message Passing Interface (MPI).
The results of test calculations are presented and an analysis of the strong and weak scalability of the developed software is carried out.

\end{abstract}

\section{Introduction}
The primary goal of any acoustic study in wells is to obtain information about the mechanical properties of adjacent rocks, such as the distribution of P and S wave velocities, density, quality factor, and others, based on the results of measuring the wave fields created by a source placed in the well. It should be emphasized that the presence of a well filled with a liquid (drilling mud) has a very significant effect on the formation of wave fields. For the first time this problem in the simplest axisymmetric formulation for a homogeneous host medium was considered in \cite{Biot_1952}. It analyzes in details the structure of the wave field caused by the action of a point source located in a well filled with a liquid, and derives dispersion relations for the specific waves that arise in this case, subsequently combined under the term "tube waves". Later this problem was considered by many authors. During the 1970s - 1980s, in a series of works by representatives of the Leningrad school (see, for example, \cite{Krauklis}), using the methods of asymptotic analysis, detailed theoretical considerations of several axial symmetric formulations were carried out, which contributed to a further deepening of understanding of the main features inherent in the flow of wave processes of acoustic study. However, this approach did not allow for obtaining an exhaustive description of wave propagation in three-dimension inhomogeneous media that arise in the study of such geological objects as zones of increased fracturing, areas of rock deconsolidation, steep faults, and so on. It is the selection of such zones and the determination of their mechanical properties that is currently of the greatest interest from a practical point of view. 
One of these problems is the considered below approach of determining the depth and degree of cement solidification in the well casing string by acoustic observations.

The advent of modern high-performance computing systems with a parallel architecture has made it possible to begin a detailed study of the full wave fields arising in the acoustic study for arbitrary 3D heterogeneous statements (\cite{PRT}). 
This problem has a number of specific features that significantly complicate the construction of efficient algorithms for its numerical solution. First of all, it has several characteristic scales. Indeed, even in the simplest case of an open well, one has to deal with two characteristic sizes - the well diameter, which is 0.1 - 0.2 m, and the maximum source-receiver distance, which can reach several kilometers. In this work, we also need to take into account the effect of the casing, that is, the presence of a steel pipe separating a well filled with liquid from the host medium. The wall thickness of such a pipe is a few centimeters and, thus, is another scale of the problem.

Another feature of the problem is the concentration of the vast majority of the wave field energy in the vicinity of the well and its almost complete absence at radial distances exceeding two or three dominant wavelengths, which does not matter for a number of basic types of waves. sources. exceed 0.5 - 0.7 meters. Thus, it is advisable to limit the radial dimensions of the computational domain to these distances. However, such a limitation should not lead to false reflections from artificially introduced boundaries, which requires the use of special boundary conditions.

To limit the computational domain, we use a perfectly matched absorbing layer (PML - Perfectly Matched Layer). Initially, it was developed for Maxwell's equations in [4], and later modified for the equations of dynamic elasticity both in Cartesian \cite{Berenger,Chew,Collino} and cylindrical coordinate systems \cite{Reshetova}. In this paper, to limit the computational domain we use a modification of the PML proposed in \cite{Reshetova} for the axially symmetric formulation

One of the most significant features of the problem of numerical modeling of wave fields in wells is the presence of an extremely contrasting cylindrical interface between the well and the surrounding medium. The description of such interfaces in the Cartesian coordinates leads to their stepped representation and, as a result, to the appearance of intense scattered waves. These waves, being an undoubted numerical artifact, may well be comparable in intensity with useful waves, and sometimes even surpass them. Therefore, we settled on using a cylindrical coordinate system to study wave processes in a well.

Thus, the main requirements for software designed to simulate numerically 3D wave propagation in wells are:
\begin{itemize}
\item Cylindrical coordinate system;
\item Effective and reliable techniques of limiting the computational domain in the radial direction;
\item Parallel software implementation.
\end{itemize}

\section{Statement of the problem}
To describe the wave propagation in the well we use the following system of partial differential equations:
\begin{equation}
\begin{array}{l}
\displaystyle
\varrho\frac{\partial\vec{u}}{\partial t}=A\frac{\partial\vec{\sigma}}{\partial r}+\frac{1}{r}\frac{\partial{\vec{\sigma}}}{\partial\phi}
+\displaystyle\frac{1}{r}\left(A-D\right)\vec\sigma+\vec{F}\left(r,\phi,z;t\right);\nonumber\\\\
\displaystyle M\frac{\partial\vec\sigma}{\partial t}=A^T\frac{\partial\vec{u}}{\partial z}+\frac{1}{r}B^T\frac{\partial\vec{u}}{\partial\phi}+C^T\frac{\partial{\vec{u}}}{\partial z}+\frac{1}{r}D^T\vec{u}+\vec{G}(r,\phi,z;t);
\end{array}
\label{eq:1}
\end{equation}
with initial conditions
\begin{equation}
\vec{u}|_{t=0}=0,\quad\vec{\sigma}|_{t=0}=0.
\label{eq:cauchy}
\end{equation}
At the interface between the liquid and the solid, that is, on the inner surface of the well, we suppose the continuity of the normal components of the velocity vector and the stress tensor: 
\begin{equation}
u_r|_{R-0}=u_r|_{R+0};
\label{eq:displ}
\end{equation}
\begin{equation}
[\sigma_{rr}|]_{r=R}=[\sigma_{rz}|{r=R}=[\sigma_{r\phi}|]_{r=R}=0.
\label{eq:contact}
\end{equation}
Matrices $A,\,M,\,B,\,C,\,D$ are as follows:
\begin{equation}
A=\left(
\begin{array}{cccccc}
1&0&0&0&0&0\\
0&0&0&1&0&0\\
0&0&0&0&0&1
\end{array}
\right);\quad
B=\left(
\begin{array}{cccccc}
0&0&0&1&0&0\\
0&1&0&0&0&0\\
0&0&0&0&1&0
\end{array}
\right);
\quad
C=\left(
\begin{array}{cccccc}
0&0&0&0&0&1\\
0&0&0&0&1&0\\
0&0&1&0&0&0
\end{array}
\right);
\nonumber
\end{equation}
\[
D=\left(
\begin{array}{ccc}
0&0&0\\
1&0&0\\
0&0&0\\
0&-1&0\\
0&0&0\\
0&0&0
\end{array}
\right)^T,\quad 
M=\displaystyle\frac{1}{\mu}\left(
\begin{array}{cccccc}
\displaystyle\frac{\lambda+\mu}{3\lambda+2\mu};&\displaystyle-\frac{\lambda}{2(3\lambda+2\mu)};&\displaystyle-\frac{\lambda}{2(3\lambda+2\mu)};&0;&0;&0\\\\
\displaystyle-\frac{\lambda}{2(3\lambda+2\mu)};&\displaystyle\frac{\lambda+2\mu}{3\lambda+2\mu};&\displaystyle-\frac{\lambda}{2(3\lambda+2\mu)};&0;&0;&0\\\\
\displaystyle-\frac{\lambda}{2(3\lambda+2\mu)};&\displaystyle-\frac{\lambda}{2(3\lambda+2\mu)};&\displaystyle\frac{\lambda+\mu}{3\lambda+2\mu};&0;&0;&0\\\\
\displaystyle 0;&0;&0;&1&;0&;0\\\\
\displaystyle 0;&0;&0;&0&;1&;0\\\\
\displaystyle 0;&0;&0;&0&;0&;1\\\\
\end{array}
\right)
\]

For the sake of simplicity, we use the same equations to describe wave processes both in an elastic medium and in a liquid. To do this, it is assumed that in the liquid the shear modulus $\mu$ becomes equal to zero, which ensures the vanishing of shear stresses and tangential components of the stress tensor.

\begin{figure}[htbp]
\begin{center}
\includegraphics[width=\linewidth]{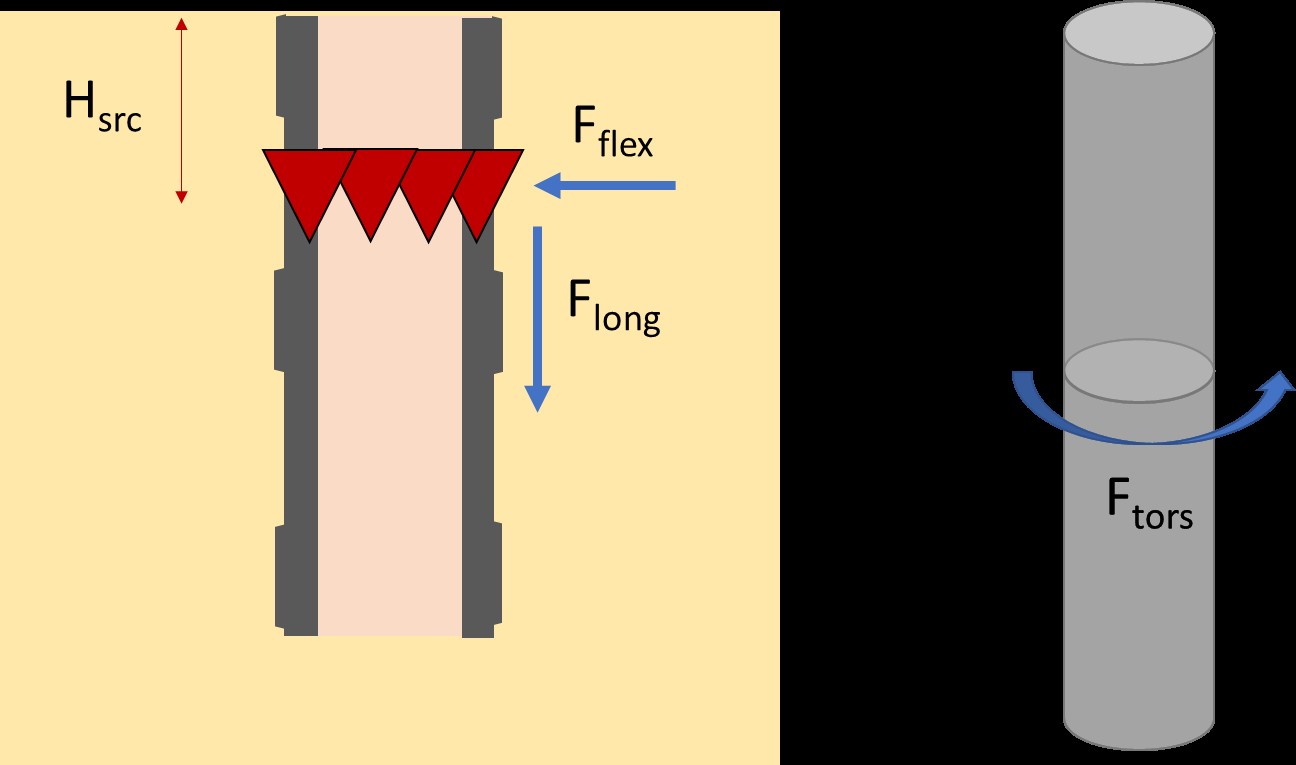}
\end{center}
\caption{The sources used for modelling. Left - point sources in horizontal ($F_{flex}$) and vertical ($F_{long}$) directions. Right - torsional sources.}
\label{Fig_sources}
\end{figure}

The following four types of sources are used to excite wave fields in a well (see Fig.\ref{Fig_sources}):
\begin{itemize}
\item Volumetric point source:
\[
\begin{array}{lcr}
\vec{F}(r,\phi,z;t)=0;\\
\vec{G(r,\phi,z;t)}=\displaystyle\frac{f(t)}{2\mu(3\lambda+2\mu)}\displaystyle\frac{\delta(r-r_0,\phi-\phi_0,z-z_0)}{2\pi r}\left(1,1,1,0,0,0\right)^T
\end{array}
\]
\item Point impact in transverse direction
\[
\begin{array}{lcr}
\vec{F}\left(\phi,r,z;t\right)=\displaystyle\frac{f(t)}{2\mu(3\lambda+2\mu)}\displaystyle\frac{\delta(r-r_0,\phi-\phi_0,z-z_0)}{2\pi r}\left(1,0,0\right)^T\\
\vec{G}\left(\phi,r,z;t\right)=0
\end{array}
\]
\item Impact of longitudinal point force:
\[
\begin{array}{lcr}
\displaystyle\vec{F}\left(\phi,r,z;t\right)=\displaystyle\frac{f(t)}{2\mu(3\lambda+2\mu)}\displaystyle\frac{\delta(r-r_0,\phi-\phi_0,z-z_0)}{2\pi r}\left(0,0,1\right)^T\\
\vec{G}\left(\phi,r,z;t\right)=0
\end{array}
\]
\item Torsion source
\[
\begin{array}{lcr}
\displaystyle\vec{F}\left(\phi,r,z;t\right)=0\\
\vec{G}\left(\phi,r,z;t\right)=\displaystyle\frac{f(t)}{2\mu\left(3\lambda+2\mu\right)}\frac{\delta\left(r-r_0,\phi-\phi_0,z-z_0\right)}{2\pi r}\left(0,0,0,1,1,0\right)^T
\end{array}
\]
\end{itemize}
As the source function f(t) we use the Ricker pulse
\[
f(t)=\left(1-2f_0^2\pi^2\left(t-\frac{2}{f_0}\right)^2\right)exp\left(-f_0^2\pi^2\left(t-\displaystyle\frac{2}{f_0}\right)^2\right)
\]
with dominant frequency $f_0$
\section{Finite-difference approximation}
\begin{figure}[htbp]
\begin{center}
\includegraphics[width=\linewidth]{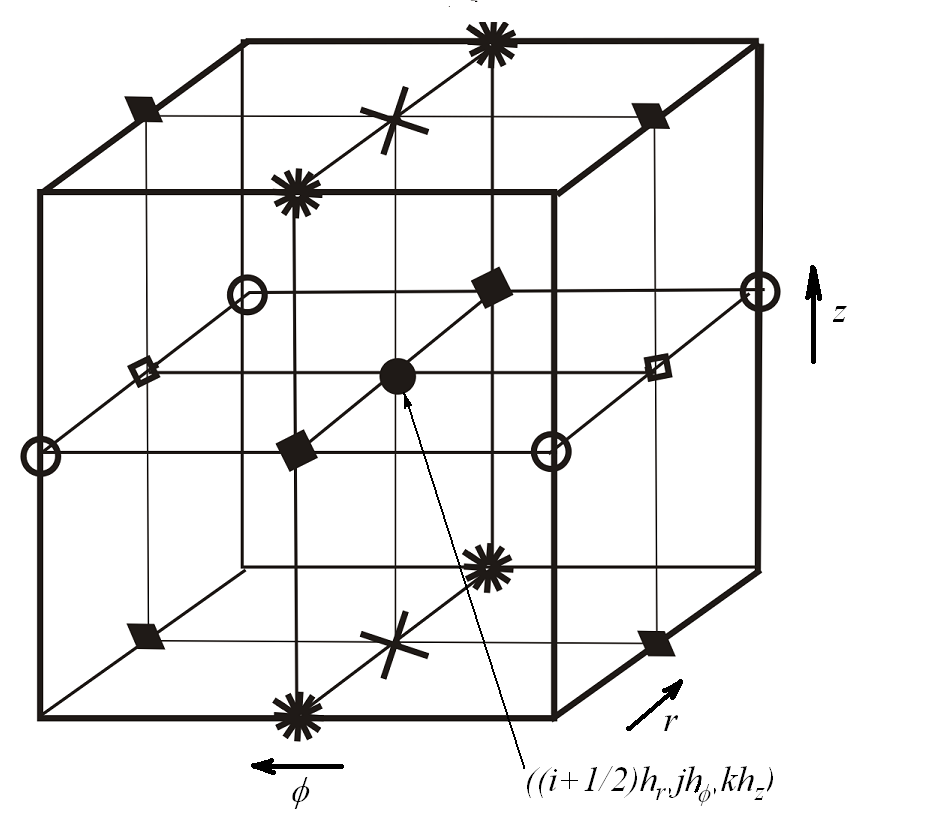}
\end{center}
\caption{Spatial FD approximation.}
\label{FD_stencil}
\end{figure}

\begin{figure}[htbp]
\begin{center}
\includegraphics[width=\linewidth]{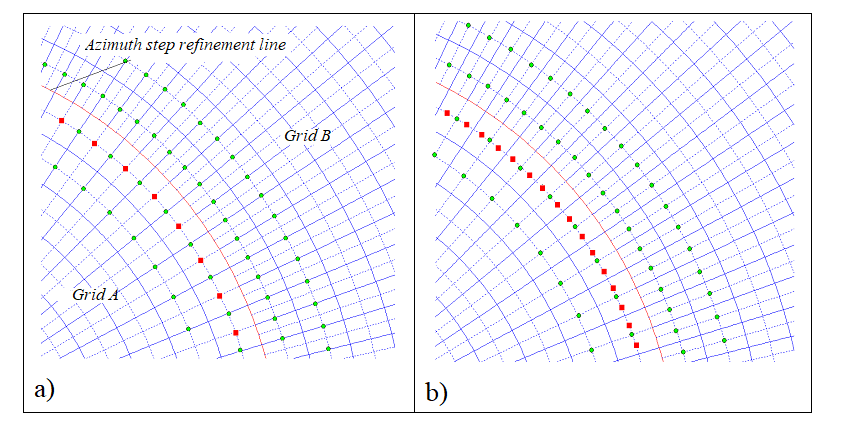}
\end{center}
\caption{Azimuth grid refinement.}
\label{refinement}
\end{figure}

For the spatial approximation of the initial-boundary value problem (\ref{eq:1}) we use the stencil presented in Fig.\ref{FD_stencil}. Elastic parameters and density of the medium are considered attached to the central point. We modified for cylindrical coordinates staggered grid finite difference scheme proposed in \cite{Virieux}. 
Let us dwell on some specific features of constructing a grid in a cylindrical coordinate system. Its characteristic feature is a linear increase in the azimuthal cell size with increasing radius r. This leads to a deterioration in the quality of the spatial approximation of the initial-boundary value problem \ref{eq:1} in the azimuthal direction at large distances from the well axis. Therefore, we produce a two-fold step reduction every time, as soon as the radius r doubles (Fig. \ref{refinement}). However, in this case, it becomes necessary to match the grids in the vicinity of a circle separating the grid with coarse (grid A) and fine (grid B) steps in the azimuth direction. It should be noted that the mutual arrangement of template nodes with integer coordinates in azimuth (Fig. \ref{refinement} a)) differs significantly from the relative position of nodes with half-integer coordinates (Fig. \ref{refinement} b). Therefore, when moving from grid A to grid B, it is necessary to interpolate the desired functions in order to determine their values at the nodes designated by rhombuses by their values given at the nodes indicated by circles. It should be emphasized that in order to preserve the second order of approximation of the original problem, such interpolation must be at least the third order of accuracy. In addition, this interpolation must be performed at each time step, which imposes very strict requirements on its performance.

The interpolation procedure based on the application of the fast Fourier transform (FFT) seems to be the most suitable in these circumstances. Indeed, the $2\pi$ periodicity in the azimuth of the desired functions ensures the exponential accuracy of this interpolation procedure, and its performance in the case when the number of nodes is equal to is beyond the competition. Based on this, we have chosen an interpolation procedure based on the application of the fast Fourier transform, performed in three steps:
\begin{enumerate}
\item Direct FFT of the initial array of elements;
\item Replenishment of the resulting array with the necessary number of zeros to the elements;
\item Inverse FFT of the augmented array.
\end{enumerate}
\section{Spatial decomposition of the target area and organization of parallel computing}
The size of the target area in the numerical simulation of wave fields in wells is tens of wavelengths in the radial direction and can significantly exceed hundreds of wavelengths in the direction along the wellbore. To ensure an acceptable level of numerical dispersion using the previously described finite difference scheme at such distances, we must take approximately 20 - 30 points per wavelength. Simple calculations show that the total amount of RAM required for this is several hundred gigabytes. Therefore, for carrying out computations, one should use high-performance computer systems with parallel architecture
\subsection{Domain decomposition}
\begin{figure}[htbp]
\begin{center}
\includegraphics[width=0.75\linewidth]{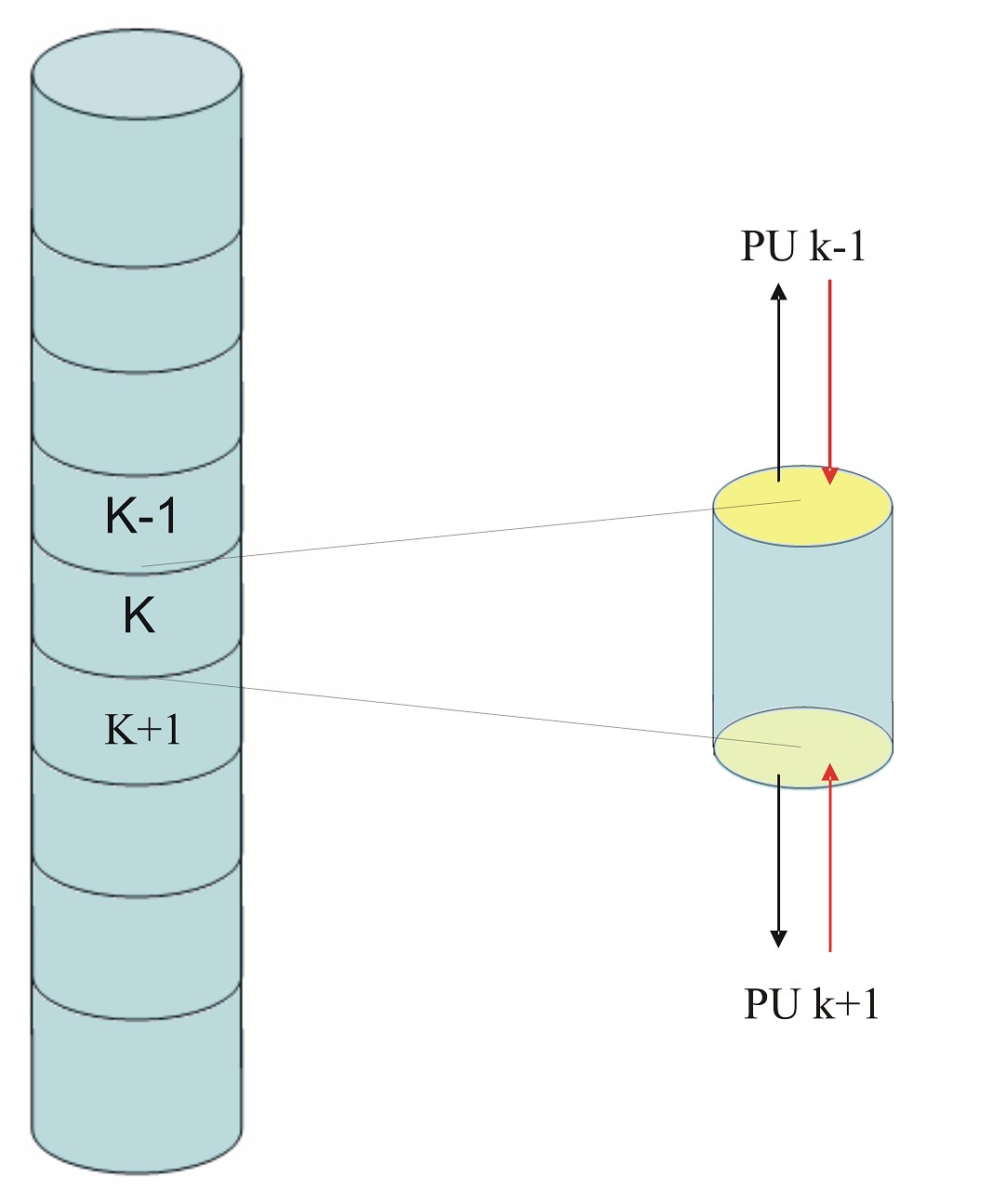}
\end{center}
\caption{Domain decomposition used for numerical simulation}
\label{Fig_DD}
\end{figure}

To manage task distribution between individual processor elements (PU - Processor Unit below) we apply the spatial domain decomposition. The specificity of the problem under consideration makes it possible to perform such a decomposition in a fairly simple and efficient way. Indeed, as noted above, the use of a cylindrical coordinate system makes it possible to choose a computational domain in the form of a cylinder, the generatrix of which is directed along the axis of a wellbore. The vertical size of such a cylinder was varying from 200 meters to 2000 meters, while its radius did not exceed half a meter. Small radial dimensions make it possible to represent the computational domain as a union of non-intersecting disks, each of which is assigned to the specific PU (see Fig.\ref{Fig_DD}). 

The explicit finite-difference scheme used requires the data exchange between neighboring processor elements. In order to understand how this interaction works, consider two adjacent disks assigned to different processor elements. They are in contact along the surface $z_k=kh_z$, and the layer $z_{k-1}$ belongs to the PU with the number $k-1$, while the layer $z_k$ already falls on the processing element with the number $k$. When moving to the next layer in time, each of these processors needs the current values of the displacement and stresses calculated by the neighbor. Therefore, before proceeding to the calculation of the desired values on the next time layer, the processor elements must exchange their current values on adjacent horizontal layers. The organization of such exchanges is carried out using the MPI  library.

Data transferring between PU brings additional overhead and slows down the computations. The computational process visualization tools currently at our disposal make it possible to obtain complete information about a load of each processor element during the execution of a particular calculation. We used the free software Jumpshot-4 to visualize the data flow during parallel computations (see https://www.mcs.anl.gov/research/projects/perfvis/software/\newline viewers/index.htm).
\begin{figure}[htbp]
\begin{center}
\includegraphics[width=\linewidth]{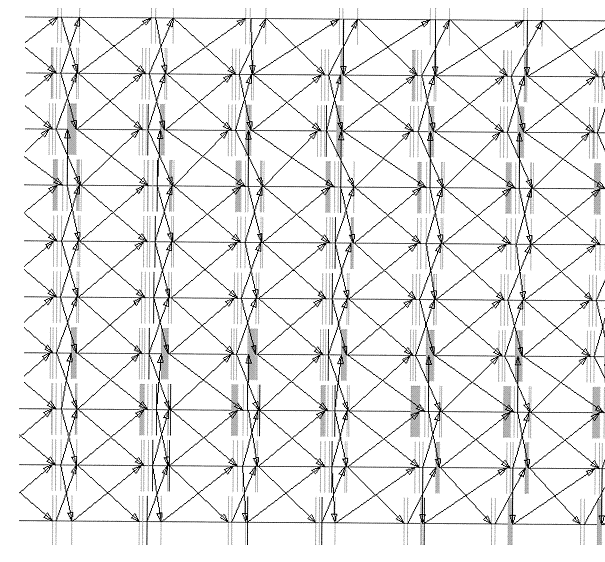}
\end{center}
\caption{Data flow and idle times for ten processor units.}
\label{Fig_Data_flow}
\end{figure}
The arrows here indicate the direction of data transfer between PUs, which are marked with vertical segments. The thickness of these segments is proportional to the idle time of a specific PU, that is, the time it takes to wait for the next amount of data from the neighbors above and below.

The total waiting time estimated using the same Jumpshot-4 software for this specific case is no more than 5\% of the total time spent on solving the problem. It is worth mentioning that the vast majority of the idle time is due to the non-uniform speed of individual PU capacity.

\subsection{Description of the hardware used}
We used the NKS-1P multiprocessor computing system (cluster) located at the Institute of Computational Mathematics and Mathematical Geophysics of the Siberian Branch of the Russian Academy of Sciences. This system consists of 27 RSC TDN421 compute modules (2xCPU Intel Xeon E5-2697Av4, 128 GB RAM), 16 RSC TSN121 compute modules (1xCPU Intel Xeon Phi 7290, 112 GB RAM), 1 RSC TDN511 compute module (2xIntel Xeon Platinum CPU 8268, 192 GB RAM), 4 RSC TDN531 compute modules (2xCPU Intel Xeon Gold 6248R, 192 GB RAM). Data exchange between parallel processes and access to the Intel Luster 200 GB parallel file system is carried out using a 100 Gbps Intel Omni-Path cluster interconnect. Peak system performance is currently 136 Tflop/s. The calculations were made using nodes on Intel Xeon E5-2697Av4 processors. A group of three nodes was used to solve each problem.
\section{Numerical experiments}
The conducted numerical experiments were aimed at solving the following problems:
\begin{itemize}
\item Determination of the possibility of detection of reflected acoustic signal from the top of cement in the annular space of the cased borehole generated at the surface;
\item Determination of parameters of acoustic source resulted in the reflected signal with maximum amplitude;
\item Determination of minimal amplitude of the source signal, so that the amplitude of the reflected signal will be above the sensitivity threshold of the measurement device;
\item Determination of the best receiver placement position on the surface part of the casing.
\end{itemize}
To obtain interesting and meaningful results from numerical modeling of the wave processes, we have started with the development of a realistic model of a well with a casing.
\subsection{Presentation of the wellbore model}
\begin{figure}[htbp]
\begin{center}
\includegraphics[width=1.\linewidth]{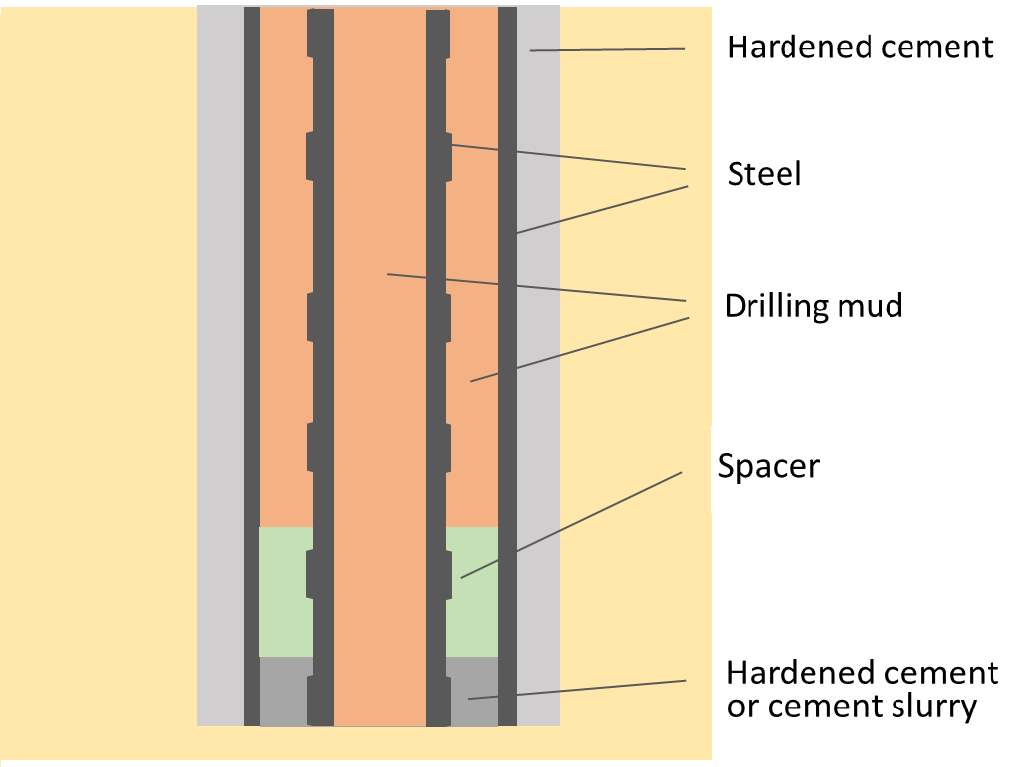}
\end{center}
\caption{The well bore model used for numerical experiments. General view}
\label{Fig_wb_model}
\end{figure}

One can see the general view of the wellbore model used for numerical experiments in Fig.\ref{Fig_wb_model}. 

\begin{figure}[htbp]
\begin{center}
\includegraphics[width=1.\linewidth]{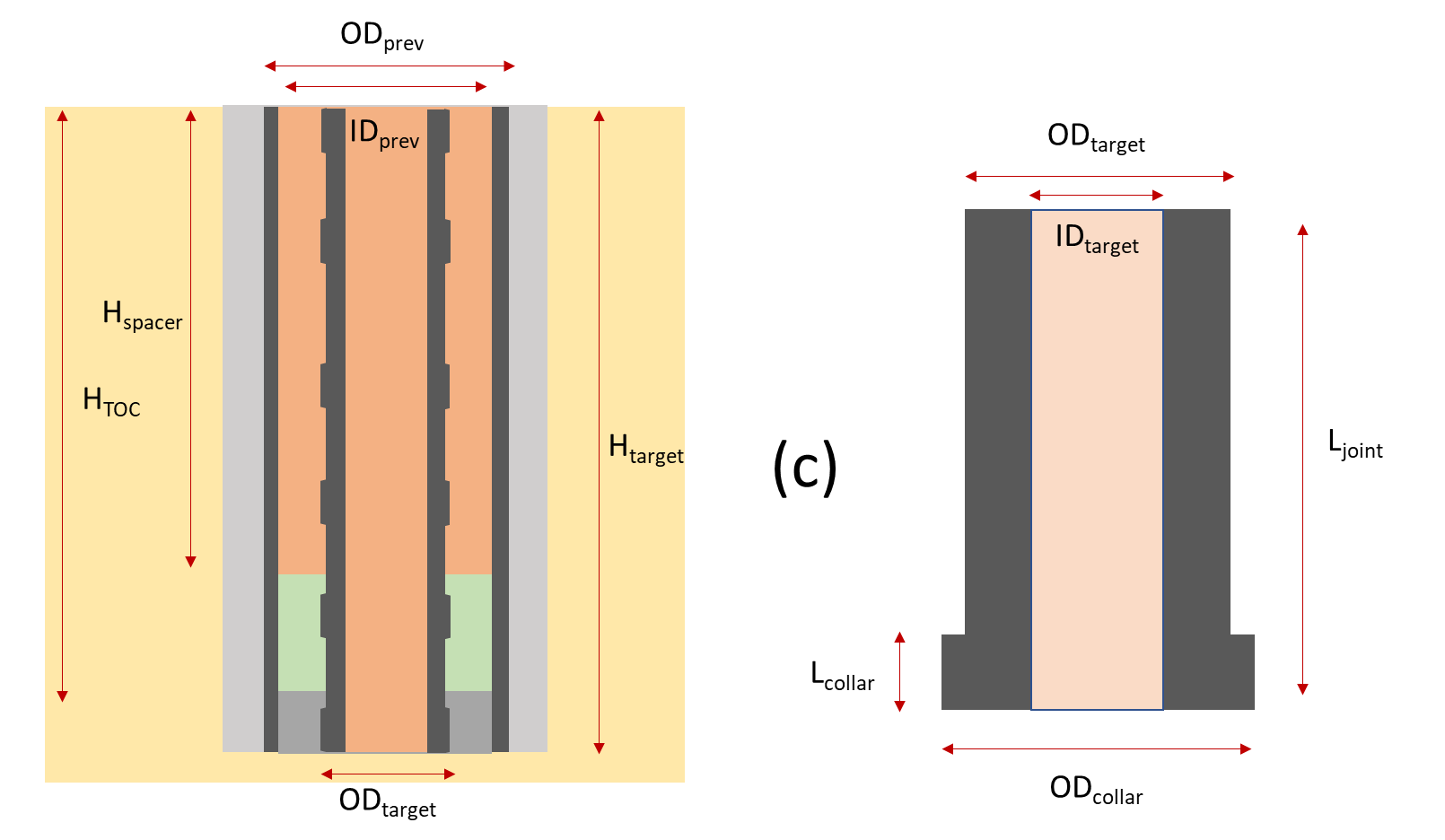}
\end{center}
\caption{The well bore model used for numerical experiments. Detailed view}
\label{Fig_wb_model_detailed}
\end{figure}

The detailed view of the model we present in Fig.\ref{Fig_wb_model_detailed} with the following values of parameters:
\newline
$H_{target}$=202 m, \,$ID_{prev}$=0.22916 m,\,$H_{toc}$=200 m,\,$H_{spacer}$=200 m, \,$L_{joint}$=12.0 m,\, $L_{collar}$=0.184 m,\,$OD_{target}$=0.1778 m, \,$OD_{collar}$=0.1945 m.

\subsection{Numerical results}
The models under consideration assume the placement of vibration sources on the column wall. Thus, in the case of a source of the "torsion action" type, torsional vibrations are excited on the outer wall of the outer column. Therefore, the waves caused by it do not propagate in the annular space, but along a metal pipe. The fact is that in the annular space, there is a drilling fluid with a zero transverse velocity, which means that a transverse wave simply cannot exist in it, that is, torsional vibrations. Therefore, the seismic wave is excited only in the outer casing and propagates along it, and outside it is absent. The speed of such a wave corresponds to the speed of a transverse wave in the metal and is equal to 3220 m/s. At the same time, it practically does not dissipate on the couplings and does not give energy to the outside. For this type of source, only the torsion component of the component is non-zero! Since seismic wave velocities for casing strings are well known and, in contrast to the drilling fluid filling the well, practically do not change, it allows  accurately estimate the depth of the hardened cement from this waves travel time.

\begin{figure}[htbp]
\begin{center}
\includegraphics[width=1.\linewidth]{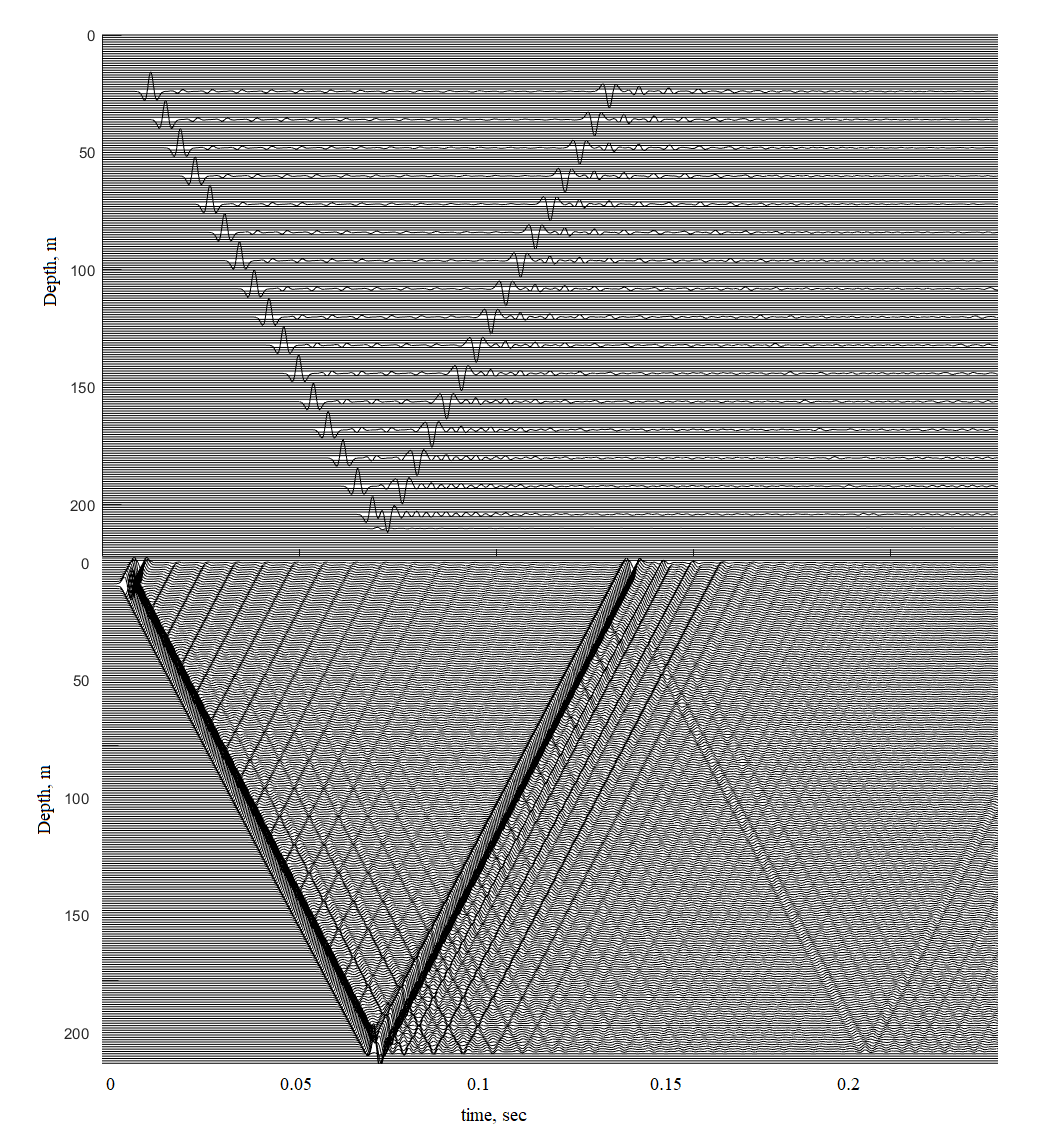}
\end{center}
\caption{The wave field by torsion source and hardened cement at the bottom. Top: recorded on the exterior casing, bottom on the interior one.}
\label{Waves by torsion source}
\end{figure}
\section{Conclusion}
Based on the application of a finite-difference approach to the approximation of the initial-boundary value problem for the system of equations of the dynamic theory of elasticity, an algorithm was developed and software focused on high-performance computing systems with a parallel architecture was created for the numerical simulation of wave fields that arise when performing acoustic logging in wells of a realistic structure. To ensure the most accurate description of the most contrasting boundary in this problem, namely, the boundary between the drilling fluid and the casing, all constructions are performed in a cylindrical coordinate system, the axis of which coincides with the axis of the well. The computational domain is limited by bordering it with the original version of a PML, the main feature of which, which distinguishes it favorably from known implementations, is the absence of splitting of variables in the azimuth direction. To organize parallel computing, the computational domain is decomposed into a number of contiguous subdomains, each of which is assigned to a separate processor element. The interaction of these processor elements is carried out using the MPI (Message Passing Interface) standard.

\section{Author contributions}
Vladimir Cheverda is responsible for the algorithm of azimuth refinement and development of PML in cylindrical coordinates and was supported by  the project 22-11-00104 of the Russian Science Foundation.\\
Galina Reshetova developed the parallel code and performed numerical experiments.\\
Artem Kabannik, Andrey Fedorov, and Roman Korkin stated the problem and developed the concept. \\
Demid Demidov, and Viktor Balalayev performed field and lab tests for the concept validation and provided data about well bore construction and parameters values.

\end{document}